\documentclass[prb,twocolumn,showpacs,amsmath,amssymb]{revtex4}

\usepackage{graphicx}
\usepackage{amssymb,amsmath}
\usepackage{dcolumn}
\usepackage{bm}
\usepackage{color}
\usepackage{siunitx} 
  
\begin{document}  
  
\title{Effective attractive polariton-polariton interaction mediated by the exciton reservoir}  
  
\author{D.~V.~Vishnevsky}  
\affiliation{Condensed Matter Physics Center (IFIMAC), Departamento de F\'isica Te\'orica de la Materia Condensada, Universidad Aut\'onoma de Madrid, 28049 Madrid, Spain}  
\email{dmitrii.vishnevsky@gmail.com}  
\author{F.~Laussy}  
\affiliation{Condensed Matter Physics Center (IFIMAC), Departamento de F\'isica Te\'orica de la Materia Condensada, Universidad Aut\'onoma de Madrid, 28049 Madrid, Spain}  
  
\begin{abstract}  
  We present a mechanism to endow the polariton gas with attractive
  interactions. The scheme relies on an exciton reservoir, which can
  be formed even if the excitons lie outside of the light-cone.  Such
  an attractive interaction should open new routes in the physics of
  quantum fluids, that we illustrate with a simple but nontrivial
  application whereby the polariton condensate enters the
  self-oscillation regime powered by the attraction from the
  reservoir. This results in pulsed emission in the GHz regime.
\end{abstract}  
  
\maketitle  
  
\section{Introduction.}  Cavity polaritons have attracted the attention
of scientists since their discovery over twenty years
ago~\cite{Weisbuch} for their unique properties as quasi-particles,
e.g., extremely small, but nonzero effective mass, interactions,
integer spin making them bosons, etc.~\cite{KavBook} This led to a
number of outstanding fundamental phenomena: Bose--Einstein
condensation (BEC), superfluidity, Josephson oscillations and many
others remarkable properties have been reported recently~\cite{BECpol,
  SFpol, JO}. One of the key features of polaritons is their nonlinear
optical properties, which opens to them the physics of interacting
systems, such as optical parametric oscillations~\cite{OPO}, bi- and
multistability effects~\cite{Baas, MSTh, MSEx}, numerous topological
defects~\cite{Vortices}, solitonic physics~\cite{Soliton}, etc.  Also,
boson-boson interaction have an important role in defining the BEC
properties. Unlike atoms which have in general attractive
interactions, which may result in a collapse of the
BEC~\cite{BECCollapse}, polaritons have repulsive
interactions~\cite{Glazov2009}. Weak attraction can occur between
cross-polarized polaritons~\cite{Vladimirova2010}, but this is in
addition to the dominant co-polarized repulsion.
  
It is increasingly appreciated that the polariton gas features one
additional strong departure from its atomic counterpart, which is the
presence of an exciton reservoir, i.e., particles uncoupled to light
that lie at higher energies and higher~$k$ vectors, and which provide
a supply of particles for the short-lived polaritons at the bottom of
their dispersion. The reservoir can extend considerably the lifetime
of a polariton experiment~\cite{Bullets} and even manifests
directly~\cite{MilenaOsc}.  As a rule, it strongly renormalizes the
interaction strength, leading to a blueshift of the polariton
emission~\cite{Gavrilov1, Gavrilov2}. The exciton reservoir arises not
only when it is created by direct incoherent excitation, but can also
be formed from resonant ground state excitation~\cite{Vishnevsky2012,
  Wouters2013}. Several experiments have reported a peculiar dynamics
of the polaritons, such as a collapse of the polariton
wavepacket~\cite{Backjet} or a negative circular polarization of the
condensate~\cite{Kulakovskii2010}. Such effects could be understood
with an effective attractive interaction~\cite{Korenev}, but its exact
nature is still unknown.
  
In this paper, we provide a mechanism to bring-in an attractive
effective interaction for the polaritons, that is dominant over the
other types of interactions. The core idea is to rely on high-$k$
excitons which indeed attract the ground state polaritons through
their excitonic component. This should open new directions both to
study the fundamental physics of these strongly correlated gas, but
also for applications since, thanks to their extremely long lifetime
and heaviness, high~$k$ excitons can provide engineerable potentials
for polaritons, which is one of the prerequisites to use them in
logical gates~\cite{OptCirc}. It seems that at the moment there is no
way to imprint a dynamical negative potential (potential wells) of
arbitrary shape. For instance, due to the repulsive
polariton-polariton interactions, optical barriers can provide only
positive potential (potential walls).  Creation of negative potentials
from the attractive interaction between cross-polarized particles has
a lot of restrictions as well, due to the instability of the
polarization and of the weakness of such an interaction. Acoustic
waves~\cite{SAW} brought up considerable tuning but do not allow for
structures of arbitrary shapes. By using the long-lived exciton
reservoir to create the potential structures, on the other hand, one
can imprint virtually any shape from the excitation spot. As one
example, we will show here how an attractive exciton reservoir can
lead to GHz self-oscillations in the emission
intensity~\cite{SelfOsc}, adding another pulsed emitter in this regime
of operation to the few known to date, e.g., with silicon photonic
crystal nanocavities~\cite{SOSN}.

\section{Polariton-polariton interactions.} Polaritons are composed from
photons and excitons and contributions from both fractions should be
considered to describe polariton-polariton interaction.  However, the
main contribution to the energy shift induced by polariton-polariton
interaction has been shown experimentally to be brought by their
exciton component~\cite{Vladimirova2010}.  Theoretical works predict
an additional term for the scattering rates due to the exciton
oscillator strength saturation that comes from the photonic
component~\cite{Glazov2009, EPL2007}, that is however small in
comparison.  It has also been shown that interactions coming from the
direct polariton-polariton scattering is of a repulsive nature, while
indirect scattering processes can be attractive, but again of a smaller magnitude. Indeed, all such scattering processes, except weak
Van-der-Waals scatterings, are for polaritons with opposite
polarization. The strength of such interactions depends strongly on
the detuning between the exciton and photon~\cite{Vladimirova2010,
  Takemura, Gavrilov2013}, and, in best case, does not exceed the
strength of the first order repulsive interaction. For this reason, we
will focus on the direct exciton-exciton interaction.

The theory was developed by Ciuti~\emph{et al.}, where the
exciton-exciton scattering was separated into three types of
processes: the direct Coulomb repulsion, the exciton exchange
interaction and the carrier (electron and hole) exchange
interaction~\cite{Ciuti1998}.  Considering two excitons with
wavevectors $\mathbf{Q}$ and $\mathbf{Q}'$ with $\mathbf{\Delta
  Q}=\mathbf{Q}-\mathbf{Q'}$, scattering each in the reciprocal space
by the amount $\pm\mathbf{q}$, the Hamiltonian describing each of
these interaction processes reads:
\begin{equation}  
{H_i^j}(\Delta Q,q,\theta ) = \frac{{{e^2}}}{{{\varepsilon _0}}}{{\frac{4}{\pi^2}}}{\lambda _{2D}}{I_i^j}(\Delta Q,q,\theta )\,,  
\label{Hamiltonian}  
\end{equation}  
where~$i,j$ label the type of interaction, $\theta$ is the angle  
between $\mathbf{\Delta Q}$ and $\mathbf{q}$, $\lambda_{2D}$ is the  
exciton Bohr radius and~$I_{i}^j(\Delta Q,q,\theta )$ some overlap  
integrals - the only part which varies with the type of
interaction. For direct and exciton exchange interactions, these
integrals admit an analytical solution  according to Ref.~\cite{Ciuti1998}:
\begin{widetext}
\begin{equation}
\begin{array}{l}
{I_{dir}}(\Delta Q,q,\theta ) = \frac{{{\pi ^3}}}{{2q{\lambda _{2D}}}}\left\{ {{{\left[ {1 + {{\left( {\frac{1}{2}{\beta _e}q{\lambda _{2D}}} \right)}^2}} \right]}^{ - 3/2}} - {{\left[ {1 + {{\left( {\frac{1}{2}{\beta _h}q{\lambda _{2D}}} \right)}^2}} \right]}^{ - 3/2}}} \right\},\\
I_{exch}^X(\Delta Q,q,\theta ) = {I_{dir}}(0,\sqrt {{{(\Delta Q)}^2} + {q^2} - 2\Delta Qq\cos \theta } ,0),
\end{array}
\label{int1}
\end{equation}
\end{widetext}
where $\beta_{e(h)}=m_{e(h)}/(m_{e(h)}+m_{h(e)})$ is the relative mass
of the electron (hole). The integral $I_{exch}^{e(h)}$ for the carrier
exchange interaction has a more complicated shape:
\begin{widetext}
\begin{equation}
\begin{array}{l}
I_{exch}^{e,h}(\Delta Q,q,\theta ) = \int\limits_0^\infty  {dx} \int\limits_0^{2\pi } {d{\theta _x}} \int\limits_0^\infty  {d{y_1}\int\limits_0^{2\pi } {d{\theta _1}\int\limits_0^\infty  {d{y_2}\int\limits_0^{2\pi } {d{\theta _2}} } } } x{y_1}{y_2}\cos \{ \Delta Q{\lambda _{2D}}\left[ {{\beta _{e,h}}x\cos (\theta  - {\theta _x}) + {\beta _{e,h}}{y_1}\cos (\theta  - {\theta _1})} \right] + \\
 + q{\lambda _{2D}}\left[ { - x\cos {\theta _x} - {\beta _{e,h}}{y_1}\cos {\theta _1} + (1 - {\beta _{e,h}}){y_2}\cos {\theta _2}} \right]\}  \times \\
 \times \exp ( - {[{({y_2}\cos {\theta _2} - {y_1}\cos {\theta _1} - x\cos {\theta _x})^2} + {({y_2}\sin {\theta _2} - {y_1}\sin {\theta _1} - x\sin {\theta _x})^2}]^{1/2}})\exp ( - x) \times \\
 \times \exp ( - {y_1})\exp ( - {y_2})\left\{ {\frac{1}{{\sqrt {y_1^2 + {x^2} + 2{y_1}x\cos ({\theta _1} - {\theta _x})} }} + \frac{1}{{\sqrt {y_2^2 + {x^2} - 2{y_2}x\cos ({\theta _2} - {\theta _x})} }} - \frac{1}{{{y_1}}} - \frac{1}{{y2}}} \right\}\,.
\end{array}
\label{int2}
\end{equation}
\end{widetext}

Generally, polariton-polariton interaction results in two processes: scattering  
of polaritons from states $\mathbf{Q}$ and $\mathbf{Q'}$ to new states  
$\mathbf{Q+q}$ and $\mathbf{Q'-q}$, and an effective renormalization  
of the polariton energy, that can be considered as a scattering from  
$(\mathbf{Q}, \mathbf{Q'})$ to $(\mathbf{Q'} , \mathbf{Q})$. The  
latter case corresponds to the matrix elements of interaction at  
$q=0$.

There are several peculiar properties of the integrals. One is that
the integral of the direct Coulomb scattering does not depend on the
difference in the excitons wavevectors $\mathbf{\Delta Q}$, but on the
scattered wave vector $q$ only. Moreover, it vanishes at zero $q$,
so the direct interaction has no influence on the states energy
renormalization. The integral of the exciton exchange interaction
$I_{exch}^X$ has also a certain symmetry with respect to $\Delta Q$
and $q$, namely, $I_{exch}^X(\Delta Q=0,q,\theta )=I_{exch}^X(\Delta
Q,q=0,\theta )$.

\begin{figure}[t]  
  \includegraphics[width=0.5\textwidth]{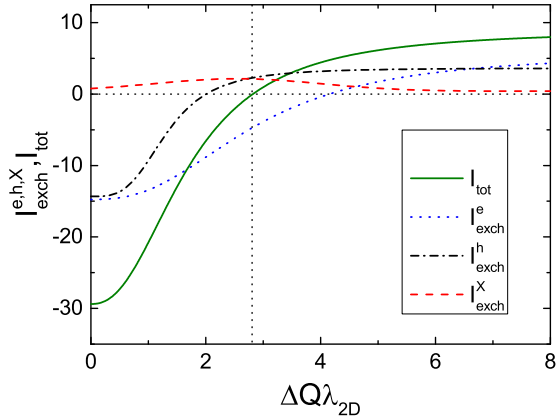}\\  
  \caption{(Color online) Integrals for electron, hole and exciton
    exchange interaction (interrupted lines), and their sum (solid
    line). The vertical dotted line separates the regions of
    repulsive and attractive interaction, providing the condition
    $\Delta Q \lambda_{2D}$ at which the character of the interaction
    reverses.}
  \label{fig1}  
\end{figure}

Figure~\ref{fig1} shows the magnitude of the  
three integrals as a function of the momentum difference $\Delta Q$  
(interrupted lines), and the total interaction (thick line). 
While the integral expressions $I_{i}^j(\Delta Q,q,\theta )$ have been  
known for over a decade and used in a large body of works~\cite{Vladimirova2010, Vishnevsky2012, Yamamoto1999}, the  
focus has been to the best of our knowledge exclusively on the  
transmitted wavevector~$\mathbf{q}$ and for zero or small exchanged  
wavevector~$\Delta Q$, since it is typically much less than the  
inverse Bohr radius of the exciton. Furthermore, the strength of the  
interaction is usually considered equal and repulsive for all  
polaritons. However, as can be seen on Fig.~\ref{fig1}, when $\Delta Q  
\lambda_{2D}>2.8$, the interaction changes its sign and becomes  
attractive. Moreover, the magnitude of the attractive interaction for  
large exchanged wavevectors is of the same order as the magnitude of  
repulsion in the common case of small exchanged wavevectors.  
  
For the usual value of a Bohr radius in the range from \num{10} to
\SI{100}{\nano\meter}, the interaction is attractive for
wavevectors difference of $\Delta Q >\SI{2.8d7}{\per\meter}$. To
provide an attractive interaction between ground-state polaritons and
high-energy excitons in GaAs structures, this value of $\Delta Q$
brings the exciton reservoir nearby the frontier of the light-cone,
and in common cases it lies outside of it. This means that these
excitons are not coupled with light, which can be advantageous to
retain them for longer times in their role of an effective reservoir,
but also impedes their introduction in the system straightforwardly in
an optical way. We therefore propose to use the polariton-polariton
scattering in order to create the high-$k$ excitons, namely, to
scatter off the Upper Polaritons Branch (UPB) with no momentum to the
lower branch, with conservation of the energy. Similar processes have
already been demonstrated~\cite{Diederichs}. To estimate the magnitude
of attraction that can thus be obtained, let us consider an ordinary
cavity with a Rabi splitting of \SI{8}{\milli\electronvolt} at zero
detuning. For a typical GaAs exciton mass of $m_X=0.5m_0$ and an
exciton Bohr radius $\lambda_{2D}=\SI{12.5}{\nano\meter}$, the
wavevector of the final state is $k_f= \SI{2.5d8}{\per\meter}$. This
corresponds to the value $\Delta Q \lambda_{2D}=3$, i.e., the
interaction between excitons and polaritons will be well into the
attractive regime.
  
\begin{figure}[t]  
  \includegraphics[width=0.5\textwidth]{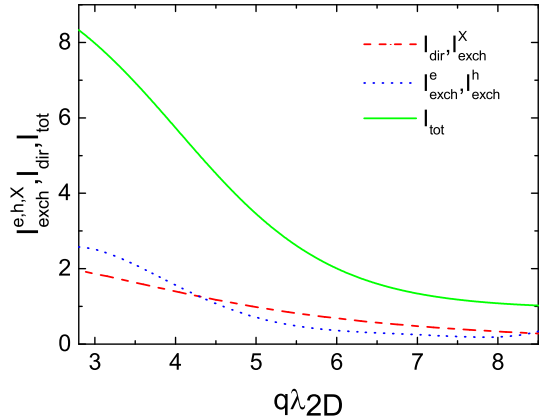}\\  
  \caption{(Color online) Integrals for electron (hole) and exciton
    exchange (direct) scattering (interrupted lines), and their sum
    (solid line). This shows that the strength of the scattering processes
    is of the same order as compared to
    the strength of polariton-polariton interaction.}
  \label{fig2}  
\end{figure}

To quantify the interaction strength to scatter polaritons from the
bottom of UPB to high-$k$ excitons, one should also calculate
Eq.~\ref{Hamiltonian} but now taking $\Delta Q=0$. In this case
$I_{exch}^e=I_{exch}^h$ and $I_{exch}^X=I_{dir}$, so only two
integrals have to be calculated. Also, since we are interested in the
scattering into attractive states, we can assume
$q\lambda_{2D}\ge2.8$. The result is presented on
Fig.~\ref{fig2}. Scattering matrix element decreases significantly as
scattered wavevector increases. Still, in the region of reciprocal
space we are interested in, $I_{tot}(\Delta Q=0,q=k_f) \approx
I_{tot}(\Delta Q=k_f,q=0)$. In other words, the strength of scattering
of two polaritons has the same order as the strength of
exciton-polariton interaction.

To describe in more details the mechanism of the reservoir formation  
under this scheme, we consider the dynamic of relaxation of the upper  
branch polaritons at~$k=0$ into reservoir excitons at~$\pm q$.  We  
must also include the thermal relaxation of these excitons to lower  
energy states, and therefore introduce a population ``$cx$'' of cold  
excitons. A set of semi-classical Boltzmann equations written for the  
second- and first-order scattering processes~\cite{CavPol} provide the  
equation of motion for their populations:  
 
\begin{equation}  
\begin{split}  
\dot{n}_{0} =& {P_0} - {\gamma _0}{n_0} + \sum\limits_k {2{w_{k0}}{n_k}{n_{ - k}}{(1 + {n_0})^2}}\\  
&- \sum\limits_k {2{w_{0k}}n_0^2(1 + {n_k})(1 + {n_{ - k}})}\,,\\  
\dot{n}_{ \pm q} =&  - \sum\limits_{cx} {{w_{qcx}}(1 + {n_{cx}}){n_{ \pm q}}} - {w_{q0}}{n_q}{n_{ - q}}{(1 + {n_0})^2}\\  
&+ {w_{0q}}n_0^2(1 + {n_q})(1 + {n_{ - q}})\,,\\  
\dot{n}_{cx} =& \sum\limits_q {{w_{qcx}}(1 + {n_{cx}}){n_q}}  - {\gamma _{cx}}{n_{cx}}\,.  
\end{split}  
\label{SCBE}  
\end{equation}  
%
We have included ${\gamma _{0,cx}}$ the polariton (exciton) inverse  
lifetime, $w_{ik}$ are the scattering rates between $i$ and $k$  
states, and $P_0$ is the pumping rate of the UPB. Since the reservoir  
excitons lie outside of the light cone, we consider that their lifetime  
is limited only by their scattering rate~$w_{qcx}$ to cold states. 
Finally, taking into account that $n_{-q}=n_{q}$, the set of  
equations is reduced to:  

\begin{equation}  
\begin{array}{*{20}{l}}  
\dot{N}_{0} = {P_0} - {\gamma _0}{N_0} - 2{W_{0R}}N_0^2,\\  
\dot{N}_{R} =  - {W_{Rcx}}{{N}_{R}} + {W_{0R}}N_0^2,\\  
\dot{N}_{cx} = {W_{Rcx}}{{N}_{R}} - {\gamma _{cx}}{N_{cx}},  
\end{array}  
\label{SCBEsimp}  
\end{equation}  
where $W_{Rcx}$ ($W_{0R}$) is the total scattering rate from the reservoir to  
cold excitons (from initial state to the reservoir), $N_{0}$, $N_R$ and $N_{cx}$ are the total densities of polaritons, reservoir and cold excitons respectively. The value of $W_{Rcx}$ is important since it constrains  
the attractive reservoir lifetime. From acoustic relaxation only, the  
effective time $\tau_{Rcx}=1/(2W_{Rcx})$ takes values from tens of  
picoseconds to nanoseconds depending on the 2D exciton Bohr radius and  
the width of the quantum well~\cite{CavPol}.    
  
For excitons with a small Bohr radius, e.g., in GaAs, although this
time, which is tens of picoseconds, is short relatively to the cold exciton radiative lifetime
$\tau_{cx}=1/(2\gamma _{cx})$, it remains long with respect to the
polariton lifetime $\tau_{0}=1/(2\gamma _{0})$, so an attractive
reservoir could be obtained for some finite duration of time, e.g., in
pulsed excitation experiments. Such conditions occur, for example, in
recent experimental work showing the collapse of the
wavepacket~\cite{Backjet}. For excitons with a large Bohr radius,
$\tau_{Rcx} \gg \tau_{cx}$ and the attractive potential can be
sustained in the CW regime. The total density of excitons in the
reservoir is easily obtained from the steady-state of the Boltzmann
equations~(\ref{SCBEsimp}):
\begin{equation}  
{N_{Rs}} = \frac{{{W_{0R}}}}{{{W_{Rcx}}}}N_{0s}^2\,,  
\label{Nrsteady}  
\end{equation}  
while for the population of cold excitons:  
\begin{equation}  
N_{cxs} = \frac{{{W_{Rcx}}}}{{{\gamma _{cx}}}}N_{Rs}=\frac{{{W_{0R}}}}{{{\gamma _{cx}}}}N_{0s}^2,  
\label{Ncxteady}  
\end{equation}  
with $N_{0s}$ the steady-state value of the polariton density:
\begin{equation}  
  {N_{0s}} = \sqrt {{{\left( {\frac{{{\gamma _{0}}}}{{4{W_{0R}}}}} \right)}^2} + \frac{P}{{{W_{0R}}}}}  - \frac{{{\gamma _{0}}}}{{4{W_{0R}}}}  
\label{n0st}  
\end{equation}  

Since $W_{0R}$ has the same order of magnitude as the strength 
of polariton-polariton interaction (it is possible to get the  
product $\hbar W_{0R}N_0$ up to \si{\milli\electronvolt} scale), and the  
reservoir exciton lifetime is very long  
($\hbar\gamma_R<\SI{0.1}{\micro\electronvolt}$), one can easily obtain the  
ratio $N_{Rs}/N_0 \gg 1$. Also, if $W_{Rcx} \ll \gamma_{cx}$, we can  
obtain $N_{Rs}/N_{cx} \gg 1$, which means that the attractive forces  
acting on the ground-state polaritons will prevail.  
  
\section{Polariton dynamics.} Equations~\ref{SCBEsimp} can now be  
extended to describe the coherent dynamics of the polaritons. We write  
a set of nonlinear Schr\"{o}dinger equations for the wavefunction of  
the upper polaritons ($\psi_{0}(\mathbf{r},t)$) coupled to the  
attractive exciton reservoir ($N_R(\mathbf{r},t)$), which, in turn,  
are coupled to the gas of repulsive cold excitons  
($N_{cx}(\mathbf{r},t)$):  
\begin{widetext}  
\begin{equation}  
\begin{array}{*{20}{l}}  
{i\frac{{d{\psi _{0}}(\mathbf{r},t)}}{{dt}} = [ - \frac{{{\hbar}}}{{2{m_{0p}}}}\Delta  - i{\gamma _{0}} + \alpha {{\left| {{\psi _{0}}(\mathbf{r},t)} \right|}^2} + 2\alpha \left| {{N_{cx}}(\mathbf{r},t)} \right|+ \beta |{N_R}(\mathbf{r},t)| - i{W_{0R}}{{\left| {{\psi _{0}}(\mathbf{r},t)} \right|}^2}]{\psi _{0}}(\mathbf{r},t) + P(\mathbf{r},t)}\,,\\  
\frac{{d{N_R}(\mathbf{r},t)}}{{dt}} = [ - \frac{{{\hbar}}}{{2{m_R}}}\Delta  - {W_{Rcx}}]{N_R}(\mathbf{r},t) + {P_R}(\mathbf{r},t)\,,\\  
\frac{{d{N_{cx}}(\mathbf{r},t)}}{{dt}} =  - {\gamma _{cx}}{N_{cx}}(\mathbf{r},t) + {W_{Rcx}}{N_R}(\mathbf{r},t)\,.  
\end{array}  
\label{NLSE}  
\end{equation}  
\end{widetext}  
  
Here, $\alpha$ and $\beta$ are the polariton-polariton and
exciton-polariton interaction constants. As discussed previously,
there are different second order processes which could affect the
value of~$\alpha$, depending on the detuning. However, $\alpha$ can
only be changed quantitavely, i.e., its absolute value can change but
not its sign. The reduction of~$\alpha$ enhances the attraction from
the reservoir, so we will consider the worst case, when $\alpha$ is
defined only by repulsive interactions. In this case, $\alpha$ is
proportional to the product $X^2I_{tot}(\Delta Q=0,q=0)$, while $\beta
\sim XI_{tot}(\Delta Q \ne 0,q=0)$. $X$ is the Hopfield coefficient
which is~$1/2$ at resonance. Taking $I_{tot}=-30$ for $\alpha$ and
$I_{tot}=5$ for $\beta$, we obtain a ratio $\alpha/\beta=-3$. The
scattering rate $W_{0R}$ is of the
same order as $\beta$. The term $P(\mathbf{r},t)$, that is responsible
for the polariton creation, describes the resonant optical injection
and is fixed externally by the experimentalist. We assumed a Gaussian
in space for the simulation. The term $P_R(\mathbf{r},t)$ describes
the exciton creation and is defined self-consistently by the equation
of motion. Its amplitude is proportional to $W_{0R}{\left|
    {{\psi_{0}(\mathbf{r},t)}} \right|^4}$, modulated by the Fourier
transform of an ellipse in the reciprocal space, as long as excitons
are created in excited states.

Excitons have a slow motion as compared to the polaritons, due to
their heaviness, and in Eqs.~(\ref{NLSE}) of the main text, we neglect indeed
the motion of the cold excitons. However, the group velocity of hot
excitons in the region of interest in the reciprocal space
($q_{R}=\Delta Q \sim\SI{e8}{\per\meter}$) is about
\SI{1}{\micro\meter} per \SI{40}{\pico\second}, which remains slow at
the scale of the polariton lifetime but should be considered at the
timescales over which the reservoir is expected to hold. Therefore, we
do not neglect the motion of the reservoir excitons in our
calculation. However, the numerical integration of Eq.~(\ref{NLSE}) in this
form, i.e., with a Laplacian for the diffusion, is costly in time. To
facilitate the computation, we approximate the reservoir exciton
motion as ballistic, i.e., we solve:
\begin{equation} 
\begin{split} 
\frac{{d{N_R}(\mathbf{r},t)}}{{dt}} = \frac{{{\hbar}}}{{2{m_R}}}q_{R}|\nabla {N_R(\mathbf{r},t)}|-\\ -{W_{Rcx}}{N_R(\mathbf{r},t)}+ {W_{0R}}{\left| {{\psi _{0}(\mathbf{r},t)}} \right|^4}.
\end{split}
\label{SERes}  
\end{equation}  

In other words, we approximate the exciton dispersion by a linear  
function, whose slope is tangent to the actual parabolic  
dispersion at high wavevectors. In this way, we can shift the excitons  
momentum to the origin of the reciprocal space without changing their  
motion. This approximation only neglects the diffusion of excitons,  
which, however, can be neglected just like the motion of the cold  
excitons. Also, for simplicity we neglect all the possible nonlinear  
terms in the equation for the exciton motion.

\begin{figure}[tb]  
  \includegraphics[width=0.5\textwidth]{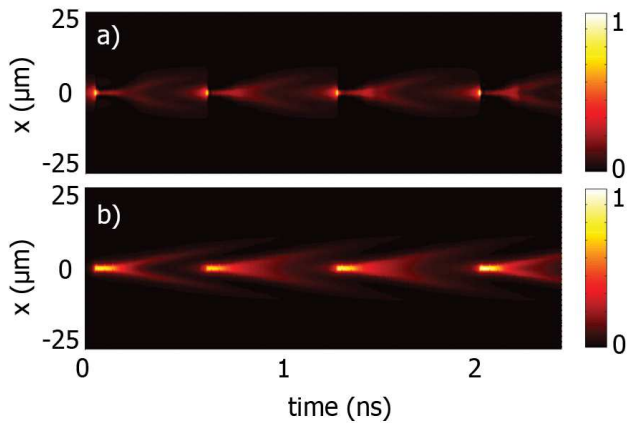}\\  
  \caption{(Color online) Time evolution of a) the polariton intensity  
    $|\psi_{0}|^2$ and b) the reservoir exciton intensity  
    $|N_R|$. The interplay of attraction and repulsion bring the  
    polariton emission into a self-oscillation regime.}  
  \label{fig3}  
\end{figure}

We can now solve Eqs.~(\ref{NLSE}) numerically. We consider a CW
coherent resonant excitation at the bottom of the UPB with a Gaussian
spatial profile in planar geometry. We take the following parameters:
$\tau_{Rcx}=\SI{2}{\nano\second}$, $\tau_{cx}=\SI{200}{\pico\second}$,
corresponding to an exciton of \SI{30}{\nano\meter} Bohr radius, and
an excitation spot of $\SI{3}{\micro\meter}$. The time evolution of
the polariton intensity is shown in Fig.~\ref{fig3}~a.  Here we have
considered the upper polariton branch only, while dealing with the
lower branch is more convenient and popular in the
laboratory due to the fast dephasing time of the UPB in most of currently available samples. Essentially the same physics and ideas apply to lower
branch polaritons in the presence of the attractive reservoir, giving
qualitatively the same behavior.

It appears right away that the interplay of attraction and repulsion
indeed results in a nontrivial dynamics of the polariton intensity,
featuring beatings in time with a period of \SI{500}{\pico\second}.
This represent a significant improvement on the recent experimental
reports of oscillating emission from a polariton
condensate~\cite{MilenaOsc}, since in our case the oscillation does
not relax towards an equilibrium or decays with the condensate, but is
sustained forever by the external continuous pumping, representing
therefore a stronger case of self-oscillations. This behavior can be
easily understood. Polaritons introduced in the cavity are partially
converted to reservoir excitons, and the competition between the
polariton-polariton repulsion and the exciton-polariton attraction
triggers a self-oscillation~\cite{SelfOsc}. When the excitation
intensity is strong enough, the attractive interaction prevails,
creating a potential dip in the center. Polaritons from the edges of
the spot start to flow towards the exciton reservoir, increasing the
number of polaritons in the center, and, therefore, also increasing
the rate of polariton-exciton conversion. The effect becomes
avalanche-like, the polariton wavepacket collapses and the number of
excitons in the reservoir increases abruptly, providing a strong
redshift for the polariton mode (up to \SI{1.5}{\milli\electronvolt}
in our calculations). The energy shift of the ground state makes the
optical injection of polaritons inefficient, resulting in a strong
diminution of the polariton intensity. The redshift is kept up to
\SI{200}{\pico\second} until the reservoir excitons flow away (their
dynamics is shown in Fig.~\ref{fig3}b), and the whole dynamics is
repeated. The relaxation of reservoir excitons is slow as compared to
their flow velocity, and a bunch of repulsive excitons, as well as the
repulsive potential that they cause, are created outside of the
excitation spot, not affecting the polariton dynamics in its center.
The transition of the system from a normal behavior with constant
emission intensity to the oscillating regime occurs when the
excitation intensity crosses a threshold $P_{th}$, which could be
roughly estimated from Eqs.~(\ref{SCBE}--\ref{Nrsteady}). The
transition between the two regimes occurs when the inward flow
compensates the outwards flow, thus establishing a steady-state in the
center of the spot. The value of the treshold~$P_{th}$ is obtained
from solving the equation that establishes the threshold $\alpha
N_{0s}+2\alpha N_{cxs}=-\beta N_{Rs}$ along with
Eqs.~(\ref{Nrsteady}), (\ref{Ncxteady}) and~(\ref{n0st}).
  
\section{Conclusions.}  We have demonstrated the possibility of providing
an effective attractive polariton-polariton interaction mediated by an
exciton reservoir. This should allow to investigate a much wider range
of dynamical behaviors and exotic phases of the polariton systems.
We simulated a simple regime of CW excitation where self-induced GHz
oscillations are sustained, showing the prospects of this mechanism
also for applications, such as clocking or pulsed lasers.
  
\emph{Acknowledgements.}  DVV would like to thank D.~Solnyshkov,  
N.~Gippius and M.~Glazov for fruitful discussions. This work was  
supported by the POLAFLOW ERC starting grant and a RyC contract.

\end{document}